\documentclass[12pt,preprint]{aastex}

\def\kann{\langle\kappa\rangle}
\def\simgt{\hbox{\rlap{\raise 0.425ex\hbox{$>$}}\lower 0.65ex\hbox{$\sim$}}}
\def\simlt{\hbox{\rlap{\raise 0.425ex\hbox{$<$}}\lower 0.65ex\hbox{$\sim$}}}
\def\hoverarrow#1{\setbox0\hbox to 0pt{\hss
                  $\scriptscriptstyle\rightharpoonup$}#1\kern.4ex
                  \raise 1.5ex\box0\kern-0.1ex}
\def\btheta{{\hoverarrow\theta}}

\begin{document}

\title{Gravitational lensing model degeneracies: Is steepness all-important?}

\author{Prasenjit Saha\altaffilmark{1,2}}
\author{Liliya L.R. Williams\altaffilmark{3}}

\altaffiltext{1}{Institute for Theoretical Physics, University of Z\"urich,
                 Winterthurerstrasse 190, 8057 Z\"urich, Switzerland}
\altaffiltext{2}{Astronomy Unit, Queen Mary and Westfield College,
                 University of London, London E1~4NS, UK}
\altaffiltext{3}{Department of Astronomy, University of Minnesota,
                 116 Church Street SE, Minneapolis, MN 55455}

\begin{abstract}
In gravitational lensing, steeper mass profiles generically produce
longer time delays but smaller magnifications, without necessarily
changing the image positions or magnification ratios between different
images.  This is well known.  We find in this paper, however, that
even if steepness is fixed, time delays can still have significant
model dependence, which we attribute to shape modeling degeneracies.  
This conclusion follows from numerical experiments
with models of 35 galaxy lenses.  We suggest that varying and twisting
ellipticities, features that are explored by pixelated lens models but
not so far by parametric models, have an important effect on time
delays.
\end{abstract}

\keywords{gravitational lensing}

\section{Introduction: why steepness?}

In the gravitational lensing of quasars by galaxies, time delays
between images are highly prized because they are proportional to the
Hubble time \citep{schech04,jakob05,kochanek06,morgan06,vuissoz06,scmw06}.
But a given set of image positions and brightness ratios---in fact any
images of sources at single redshift---can be produced by very
different lensing-mass distributions.  In particular, making a lens
profile steeper lengthens the time delays and reduces the overall
magnification, but otherwise has little or no effect on the images.

A more precise version of the previous statement is that replacing
$(1-\kappa)$ everywhere on a lens by $\lambda(1-\kappa)$---where
$\kappa$ is the projected density in units of the critical density and
$\lambda$ is a constant---multiplies all time delays by $\lambda$ and
multiplies all magnifications by $\lambda^{-2}$, but changes nothing
else.  In fact the transformation only needs to be applied within a
circle larger than all the images.  The simplest interpretation is a
stretching of the arrival-time surface by a factor of $\lambda$ along
the time axis.
Multiplying $(1-\kappa)$ by a
constant naturally makes the mass profile steeper or shallower.  That is
not exactly the same as changing the radial index, but quite similar to
it over the scales of interest.

This degeneracy has a long history and several names, having been
independently discovered at least four times.  \cite{fgs85} derived it
as a consequence of the lens equation, and the same authors in
\cite{gfs88} named it the `magnification transformation'.
\cite{paczynski86} discovered it in the context of microlensing.
\cite{ss95} found it in cluster lensing and called it a `global
invariance transformation'. \cite{wp94} came upon it as a parameter
degeneracy in galaxy-lens models.  Nowadays the common name is
`mass-sheet degeneracy'; ADS first shows the phrase used by \cite{bn95},
but it seems the name was already in spoken usage by then. Unfortunately, the
name `mass-sheet degeneracy' can give the incorrect impression that
simply adding/removing a mass sheet is a degeneracy.  It seems preferable
to use the more descriptive term {\em steepness degeneracy,} thus avoiding
the possible confusion.  In this paper we will use `steepness degeneracy'
in both strict and rough senses: the strict meaning being rescaling
$(1-\kappa)$ within a circle enclosing all the images, and the rough
meaning being changing the radial index.

Whatever the name, the steepness degeneracy has been much discussed in
recent years \citep{bradac04,schech04,treu04,oguri02,wuck02}.  
On the other hand, there has been
little research on whether any other degeneracies are important for the
time-delay problem.  Several known lensing degeneracies are summarized
in \cite{saha00}, along with a derivation of the arrival-time interpretation
above, but apart from steepness and the obvious monopole
degeneracy, none of them are applicable in the context of lensed
quasars.

It is easy to imagine further degeneracies: we can simply make the
stretching factor a function of position.  In other words we replace
the arrival-time surface $\tau(\btheta)$ by
\begin{equation}
  \tau'(\btheta) = \lambda(\btheta) \tau(\btheta) .
\label{trans}
\end{equation}
We must require $\nabla\lambda=0$ at the image positions to preserve
said image positions, $\nabla(\lambda\tau)\neq0$ except at the images
so as not to introduce new images, and $\nabla^2(\lambda\tau)\geq0$
everywhere to keep the density non-negative.  But otherwise the
transformation (\ref{trans}) are arbitrary.
We may call such transformations {\em shape
degeneracies,} because they change the shape of the arrival-time
surface and the mass profile in some complicated way.  General shape
degeneracies change magnification ratios between different images and
time-delay ratios between different pairs of images, though particular
shape degeneracies may preserve some or all of these. In contrast, the
steepness degeneracy preserves all time-delay ratios and magnification
ratios.  Hence the effect of steepness degeneracies will be reduced if
such data are present.  If sources at multiple redshifts are present,
then steepness degeneracy is broken, while shape degeneracies can be
greatly reduced.

The only explicit example of a shape degeneracy in the literature is a
special but intriguing model constructed by \cite{zq03}, to which we
will return later.  The main aim of this paper, however, is to assess
whether shape degeneracies are important in galaxy lenses independently
of particular examples.  We can do so using pixelated modeling, which is
the best available way to explore the full range of shape degeneracies
because shape degeneracies are generically present in free-form lens models.  
(Parametric modeling, on the other hand, allows only for a restricted set
or sets of shape degeneracies.)  The trick is to somehow `turn off' the 
steepness degeneracy, and then see how degenerate time delays remain.

\section{Numerical experiments with lens models}\label{exper}

The {\em PixeLens\/} code \citep{sw04} is particularly well-suited to
exploring a large variety of models, because it can automatically
generate ensembles of models constrained to reproduce observed image
positions, and also observed time delays and tensor magnifications if
available.  The models are also constrained by a prior reflecting
conservative assumptions about what galaxy mass profiles can be
like.\footnote{We do not have dynamical models for the lenses, in the
sense of phase-space distribution functions that self-consistently
generate the three-dimensional gravitational potential.  Models of
this type are commonly fitted to stellar-dynamical data
\citep{bender05,capellari06}.  But getting the stellar dynamics
self-consistent while also fitting the lensing data has not yet been
attempted.} Details and justification of the prior are given in the
earlier paper, but basically the mass maps must be non-negative and
centrally concentrated with a projected radial profile steeper than
$R^{-0.5}$.

In {\em PixeLens\/} it is easy to turn off the steepness
degeneracy: we can simply constrain the `annular density' $\kann$,
meaning the average $\kappa$ in an annulus between the innermost and
outermost images, to some pre-specified value. Since $\kann$ is linear
in the mass profile, it is easily incorporated by {\em PixeLens\/} as
an additional constraint.  Doing so naturally blocks any global
rescaling of $(1-\kappa)$.

That $\kann$ is strongly coupled to the steepness degeneracy was
pointed out by \cite{kochanek02}, who derived the relation
\begin{equation}
H_0 = A(1-\kann) + B\kann\alpha + C + O\left((\Delta R/R)^2\right) \; .
\label{csk1}
\end{equation}
for lens models with given image positions and time delays.  Here
$\alpha$ is the radial index as in $\kappa\sim R^{-\alpha}$, $A,B,C$ 
are constant for any given lens system, and $\Delta R/R$ expresses the 
thickness of the image annulus. The $A$ coefficient is, roughly speaking, 
the highest $H_0$ allowed by a given set of image positions and time 
delays. If steepness dominates, then $B,C,$ and the error term will be 
small. A test of Eq.~(\ref{csk1}) for pixelated models of six time-delay 
lenses has already been presented in \cite{sw04} (Figs.~11 and 14).  In 
order to test it also for lenses without measured time-delays, it is 
convenient to rewrite (\ref{csk1}) in dimensionless form, which we now do.

Consider the scaled time delay for a given lens defined by
\begin{equation}
\varphi = {16\over(R_1+R_2)^2\,D}\,H_0\,\Delta t ,
\end{equation}
where $\Delta t$ is the time delay between the first and last images
in arrival-time order, $R_1,R_2$ are the lens-centric sky distances of
the same images, and $D$ is the dimensionless cosmology-dependent
factor $(1+z_{\rm L})(H_0/c)D_{\rm L}D_{\rm S}/D_{\rm LS}$.  The factor
$\frac1{16}(R_1+R_2)^2$ in steradians is roughly the fraction of the
sky covered by the lens, and it turns out to be of the same order as
$H_0\Delta t$.  In other words, the sky-fraction of the lens is
roughly the time delay divided by the Hubble time \citep{saha04}. The
scaled time delay $\varphi$ ranges from 0 to about 8, and correlates
with the image morphology.  We will see this in detail later.

Multiplying Eq.~(\ref{csk1}) by $16\Delta t/(R_1+R_2)^2D$ gives
the dimensionless relation
\begin{equation}
\varphi = a(1-\kann) + b\kann\alpha + c + O\left((\Delta R/R)^2\right) \; .
\label{csk2}
\end{equation}
with new constants $a,b,c$ proportional to $A,B,C$.  $H_0$ is now
eliminated.  If we now examine the model-dependence of $\varphi$ at
fixed $\kann$ for any lens, we will have the size of the error term,
or alternatively the contribution of degeneracies not considered in
Kochanek's derivation.

To investigate the model-dependence of $\varphi$ we considered 35
galaxy lenses in three modeling stages. The purpose of the first stage
is to `fill in' the information gaps in the observed lensing data, mostly
time delays, with plausible values.\footnote{We do not claim that the
time delays we generate are accurate estimates of the actual time 
delays---for the purposes of this paper it is adequate to use 
reasonable values.} The models resulting from the second stage modeling 
allow for both the steepness and shape degeneracies. But the models of
the third stage have the steepness degeneracy suppressed, leaving shape
degeneracies only.

In the first modeling stage, we generated ensembles of 200 models for
all 35 lenses, using image positions, plus time delays if available,
and imposing $H_0^{-1}=14\,\rm Gyr$.  The image positions were taken
from the CASTLES compilation \citep{castles} in most
cases.\footnote{We tried to include all the well-studied lenses, but
omitted the `cloverleaf' H1413+117 because there seems to be a
significant uncertainty in the galaxy position. In such a highly
symmetric system, an uncertain lens center position causes ambiguity
in the time-ordering of images, which is fundamental to our modeling
technique.}  For one lens, J0414+053, we specified three VLBI
components \citep{twh00} as distinct image systems, thereby
constraining the relative tensor magnifications.  In 27 of the lenses
we required the models to have inversion symmetry.  In 8 lenses we let
the models be asymmetric, either because secondary lensing galaxies
have been identified or because symmetric and asymmetric assumptions
led to very different mass distributions.  Earlier blind tests
\citep{ws00} indicate that the latter procedure is quite successful at
identifying asymmetric lenses.

\begin{figure}
\epsscale{0.6}
\plotone{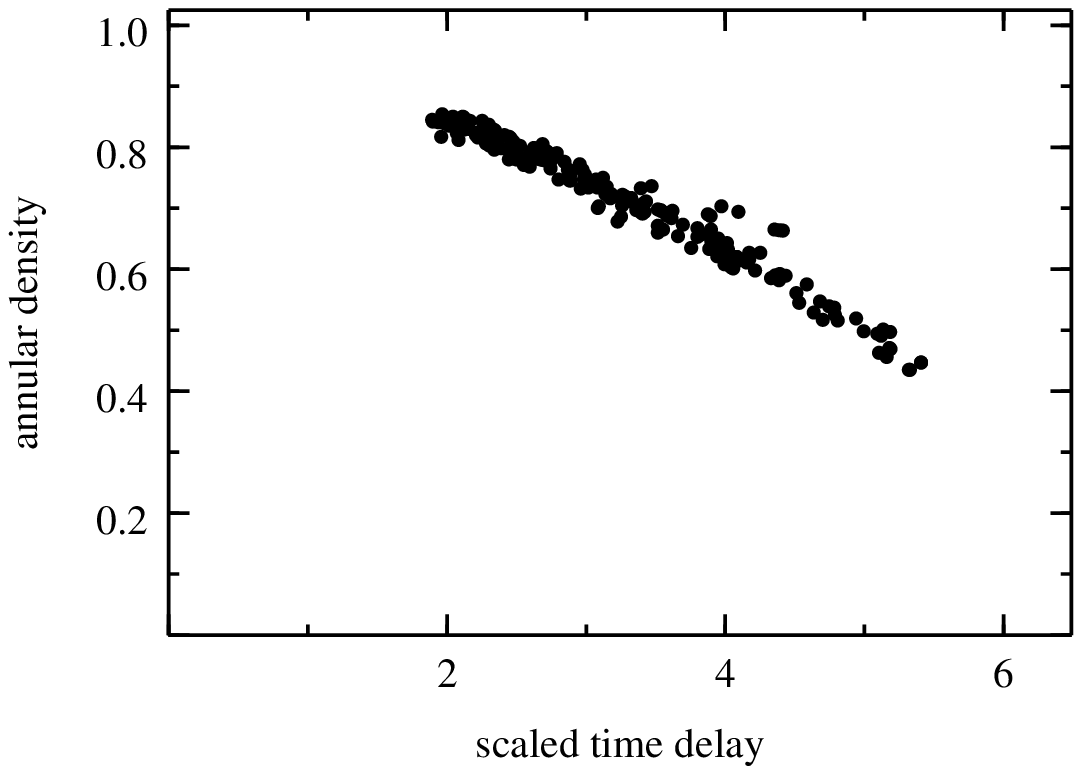}
\caption{Plot of $\kann$ against $\varphi$ for an ensemble of 200
models of B1422+231, using the observed image positions and some
plausible time delays.}
\label{corr1}
\end{figure}

In the second modeling stage, we used the ensemble-average values from
the first stage to fill in all unmeasured time delays.  Then we
removed the constraint on $H_0$, and generated model ensembles again.
In second-stage models, all members of a model-ensemble for a given
lens have the same image positions and time delays, but $\varphi$ and
$\kann$ vary. Fig.~(\ref{corr1}) shows the variation of $\varphi$ with
$\kann$ in second-stage models for the long-axis quad\footnote{We will
use the names core quad, inclined quad, long- and short-axis quad,
axial double, and inclined double to describe image morphologies. See
\cite{sw03} for details.} B1422+231. Clearly $\varphi$ is nearly
linear in $\kann$, and moreover the intercept on the $\kann$ axis is
close to $\kann=1$, hence $a(1-\kann)$ is a good fit.  The dispersion
in $\varphi$ is $\sim25\%$.\footnote{By fractional dispersion we mean
$\frac12({\rm 84th\ percentile}-{\rm 16th\ percentile})/{\rm median}$.
For a Gaussian, that would be $\sigma/m$.}

\begin{figure}
\epsscale{0.6}
\plotone{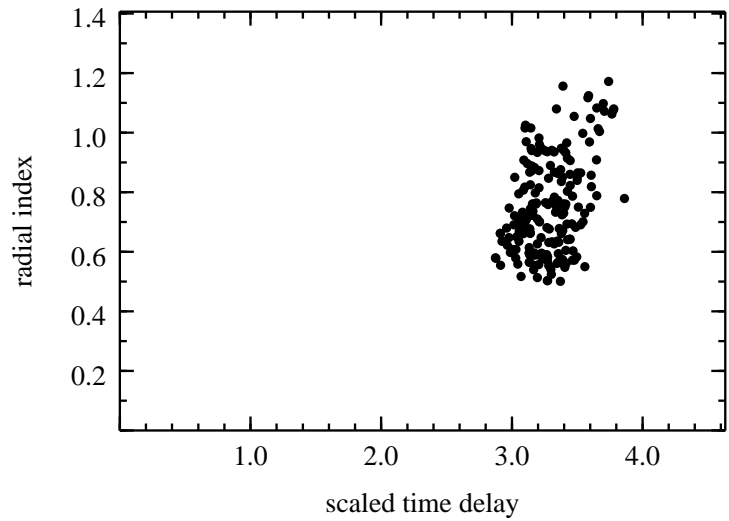}
\caption{Plot of $\alpha$ against $\varphi$ for 200 models of
B1422+231, with $\kann$ fixed at the ensemble average from
Fig.~\ref{corr1}.}
\label{corr2}
\end{figure}

For the third modeling stage, we constrained $\kann$ to its average
value for first-stage models.  Thus, all third-stage models of a lens
have their time-delays and $\kann$ fixed at either the measured or
some plausible value, thus suppressing the steepness degeneracy, while
the variation of $\varphi$ charts the $b$, $c$ and error terms in
Eq.~(\ref{csk2}). Fig.~(\ref{corr2}) shows this variation for
B1422+231 again. A small positive $b$ coefficient ($b\approx a/10$) is
noticeable, but is largely drowned out by variation from other
degeneracies.  Clearly, if steepness is the dominant degeneracy, as is
the case with B1422+231, the correction terms given by Kochanek ($B$
and $C$, and the error term) provide little improvement.

Detailed results from the third-stage modeling, i.e., with steepness
degeneracy turned off, are shown in Figs.~\ref{map1}--\ref{map4}.
These figures show the $\Delta\varphi/\varphi$ (meaning the fractional
dispersion of $\varphi$ in third-stage models) against the mean
$\varphi$ for all 35 lenses, using mass maps of the lenses themselves
as plotting symbols. Figs.~\ref{map1}--\ref{map3} should be considered
overlaid, while Fig.~\ref{map4}, containing the highly asymmetric
lenses, uses a different scale. The dispersion $\Delta\varphi/\varphi$
quantifies the relative effects of the steepness and shape
degeneracies. Systems where steepness dominates have small
$\Delta\varphi/\varphi$, for example 4\% in the case of B1422+231, while
systems where shape degeneracies dominate have considerably larger
$\Delta\varphi/\varphi$, $\simlt$ 40\%.

The immediately striking conclusion is that although in some lenses
(including B1422+231) the time delay variation is dominated by the
steepness degeneracy, in general shape degeneracies are important.

Could this result be an artifact of the pixelated method?  We must
consider the possibility that the ensembles contain models with
irregular structures not present in real galaxies, because irregular
structures would tend to get washed out in ensemble averages while
still contributing a large scatter to $\Delta\varphi$.  We can
spot-check for this possibility by inspecting individual models from
the ensembles.  In Figs.~\ref{isol1} and \ref{isol2} we do so for
B1422+231 and J1411+521 respectively.  B1422+231 is an axial quad, as
we have already noted, and has $\Delta\varphi/\varphi\simeq4\%$, while
J1411+521 is a core quad with $\Delta\varphi/\varphi\simeq20\%$.  For
each of these lenses, we arbitrarily select model no.~100 out of the
ensemble of 200, and show its mass profile, lens potential, and
arrival-time surface.  Comparing the two mass maps with the
corresponding ensemble-average mass maps shown in miniature in
Figs.~\ref{map1} and \ref{map2}, it is clear that ensemble averages
smooth out pixel-to-pixel variation.  But such variation affects only
the second derivative of the lens potential; the potential itself is
always smooth, as these figures show.  Furthermore, the arrival-time
contours show no spurious extra images.  When we examine many more
individual models spurious images do sometimes appear, but rarely
(perhaps 10\% of models).  The remaining noticeable difference between
the sample and ensemble-average maps is varying ellipticity,
especially the twisting ellipticity in Fig.~\ref{isol2} for J1411+521.
Roughly speaking, the sample model for J1411+521 suggests a bar but
the ensemble as a whole does not.

We can further test whether our models are exaggerating the scatter in
time delays by comparing with Table~2 in \cite{kochanek02}.  The table
shows that (a)~for the axial doubles 1520+530, 1600+434, 2149-274, the
approximation $H_0\simeq A(1-\kann)+B\kann\alpha+C$ comes to within
$\sim5\%$ of a full model, and is a slight improvement on the lowest
order approximation $H_0\simeq A(1-\kann)$, while (b)~for the inclined
quad B1115+080, the simpler approximation comes within about $15\%$ of
a full model, and introducing $B,C$ makes the approximation worse.
The $\Delta\varphi/\varphi$ that we compute are very consistent with
these levels.  In other words, for these lenses pixelated models give
a similar estimate for the size of the error term in Eq.~(\ref{csk2})
as do Kochanek's original parameterized models.

We thus conclude that the identification by Kochanek of $\kann$ as a
tracker of the steepness degeneracy was an important insight, but the
attempt to improve beyond $H_0\simeq(1-\kann)$ had limited success
because correction term(s) due to shape degeneracies are not a function 
of $\kann$.  Consequently, the error term in Eq.~(\ref{csk2}) is not
in practice a negligible effect: on the one hand $\Delta R/R$ is not
$\ll1$ except in core quads; on the other hand, in core quads
$\varphi$ is itself small, and hence small changes in the mass profile
can produce large fractional changes in $\varphi$.  Furthermore, the
possibility of shape degeneracies of order $\Delta R/R$ (i.e., lower
order than the error term) is not ruled out.

Returning to Figs.~\ref{map1}--\ref{map4} and examining them in more
detail, we see that both $\varphi$ and its dispersion depend on the
morphology, but in different ways.  The time delay increases with
morphology as follows:
\begin{enumerate}
\item core quads ($\varphi\leq1.5$),
\item inclined quads ($1.5\leq\varphi\leq2$),
\item axial quads ($2\leq\varphi\leq4$),
\item doubles ($3\leq\varphi\leq8$).
\end{enumerate}
The relation of $\varphi$ to the morphology of the image distribution
in the lens is discussed in \citet{saha04}.

The total dispersion in $\varphi$ {\em without\/} constraining $\kann$ is
of order 25\% for all morphologies, though we have only shown
B1422+231 here.  But if $\kann$ is constrained, thus pegging the
steepness degeneracy, the residual variation in time delays increases
not like $\varphi$, but as follows:
\begin{enumerate}
\item axial systems, whether doubles or quads have
$\Delta\varphi/\varphi\sim$ 5--15\%,
\item inclined systems have $\Delta\varphi/\varphi\sim$ 5--20\%,
\item core quads $\Delta\varphi/\varphi\sim$ 5--20\%,
\item and strongly asymmetric lenses have $\Delta\varphi/\varphi\sim$ of 25\% or more.
\end{enumerate}
$\Delta\varphi/\varphi$ tracks the relative contribution of shape degeneracies.
Perhaps not surprisingly, shape degeneracies are most important in asymmetric
lenses.

\section{Discussion}\label{disc}

The steepness degeneracy in lensing is now well understood.  The above
numerical experiments attempt to estimate the effect of other
degeneracies.  This is done by searching through mass models at fixed
image-positions, time-delay ratios (where applicable), and mean
annular density $\kann$.  The additional degeneracies, quantified
approximately by $\Delta\varphi/\varphi$ at fixed $\kann$, turn out
for some lenses to be as important as steepness.

What then are the additional important degeneracies beyond steepness?
Do common parametric forms for lenses already allow for the other
degeneracies, and if not, what new parameters are needed?
Detailed answers to these questions require more research, but we can
deduce partial answers by thinking about the arrival-time surface.
In the Introduction we classified degeneracies into steepness and
shape, with the stipulation that the latter category can be further
subdivided depending on how many image observables we care to consider.
In this Section we go a little further and attempt a more quantitative,
but still intuitive classification.

Recall that the steepness degeneracy amounts to a homogeneous
stretching or shrinking of the time scale in the arrival-time surface.
Imagine now that we stretch the time scale on the E side and shrink it
on the W side, preserving the image positions. No change is required
in the circularly averaged $\kann$. The resulting models are not 
steepness-degenerate, but the time delay between E and W images will change,
producing a shape-degeneracy transformation. This particular kind
is allowed only in asymmetric lenses, but there it
may well be as important as the steepness degeneracy.  Next, let us
imagine stretching the time scale on the E and W quadrants while
shrinking it on the N and S quadrants.  Such a transformation, allowed
in inversion symmetric lenses, is likely to most affect core quads,
and inclined quads and doubles to a lesser extent, but not axial
systems.  Further, we can imagine a transformation that shrinks the
time scale at small radii and stretches it at large radii.

We can thus imagine a hierarchy of lensing degeneracies, from an $m=0$
mode (the steepness degeneracy) through $m=1,2,$ etc.\ representing various
shape degeneracies. This is reminiscent of basis functions in cylindrical
coordinates, but we emphasize that shape degeneracies are not
additive modes in the arrival-time surface,
still less so in the mass profile --- they are multiplicative modes in
the arrival-time surface, and in the mass profile their form will be
more complicated.

The steepness degeneracy is special in that it rescales the arrival
time surface homogeneously, leaving time-delay ratios and
magnification ratios unaffected, while there is no guarantee that
shape degeneracies will preserve time-delay and magnification ratios. 
The image elongation information by itself, as measured in weak lensing 
does not break shape degeneracies, but having having many weakly lensed
images would help to constrain the shape of the arrival time surface.  
Sources at multiple redshifts will break steepness, and help reduce
shape degeneracy.

We can try and guess the sort of mass-profile feature that
will produce an $m=2$ mode.  By analogy with the steepness degeneracy,
suppose an elliptical mass profile is steeper along the long axis than
the short axis; this corresponds to ellipticity decreasing with
increasing radius, and it seems plausible that it will increase time
delays along the long-axis direction and decrease delays along the
short-axis direction.  In general we suggest that ellipticity varying or
twisting with radius as the signature of $m=2$ and higher modes.
Re-examining our early models of the inclined quad B1115+080
\citep{sw97} the role of such features in fitting time delays is already
apparent; at the time we commented briefly on it but had no
interpretation.

The above suggests interpreting the degeneracy given by \cite{zq03} as
a mixture of steepness and shape degeneracies.  Their Fig.~2 illustrates 
the transformation of an arrival-time surface, which appears to be an 
$m=2$ stretching/shrinking followed by an $m=0$ stretching with the 
effects canceling at the image positions but not globally.  (Note that 
the left- and right-hand sides of their arrival-time plot actually 
correspond to a $90^\circ$ change of position angle, not $180^\circ$.)

In the Zhao-Qin example, the ellipticity in the potential comes
entirely from external shear and the main lens is circular.  But in
our Figs.~\ref{map1}--\ref{map4}, varying and twisting ellipticity is
a common feature, especially in inclined systems.  The axial systems
in these figures tend not to show twisting ellipticity.  Recall also
from our numerical results that axial systems like B1422+231 tend to
have the lowest $\Delta\varphi/\varphi$, that is to say, steepness
dominates.  Individual models of axial systems may still contain
twisting ellipticity; however, clockwise and anti-clockwise twists are
equivalent if the image morphology is axial, hence such twists will
tend to cancel in the ensemble average. For inclined image
morphologies, clockwise and anti-clockwise twists in the density are
not equivalent, and will tend to survive in an ensemble average.  We
may ask whether the pixelated method tends to exaggerate twisting
ellipticity.  The blind tests in \cite{ws00} are reassuring in this
regard; no spurious twisting appears in the ensemble-average models.

We remark that in galaxy dynamics, twisting ellipticity arises
naturally in at least two ways: differential rotation leading to
spiral features, and projection of triaxial features. Because these and
other shape features can be important in real lenses, the errors in
derived $H_0$ must incorporate all of the degeneracies\citep{scmw06}.

The arguments in this Discussion are hand-waving, but they indicate
that the issue of
varying/twisting ellipticity needs closer attention.  One project that
is now called-for is to map the degeneracies in pixelated models in
detail, using principal components analysis or similar on model
ensembles, to see if a hierarchy of degeneracies indeed emerges.
Another project is to incorporate ellipticities that can vary or twist
with radius into parametric models.



\begin{figure}
\epsscale{1}
\plotone{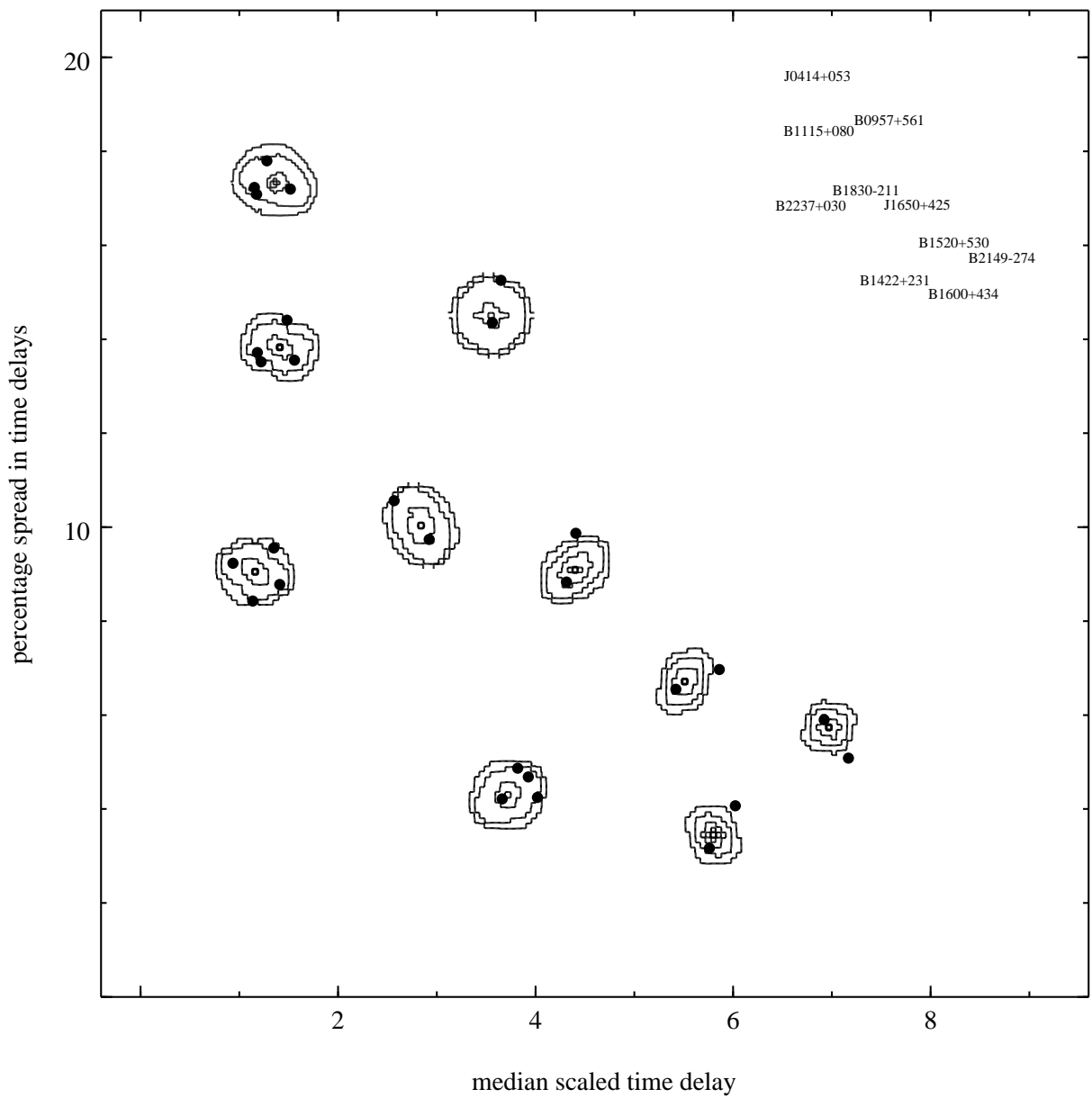}
\caption{Plot of the dispersion $\Delta\varphi/\varphi$ against median 
$\varphi$ at fixed $\kann$. Each plotting symbol is the ensemble-average mass
map of the lens, with the image-positions indicated.  The mass
contours are in logarithmic steps of $10^{0.4}$ (like a magnitude
scale) and the third contour from the outside is always $\kappa=1$.
But note that the spatial scale is different for different
lenses. Lens names on the upper right form a key.  All models except
J0414+053 have inversion symmetry.}
\label{map1}
\end{figure}

\begin{figure}
\epsscale{1}
\plotone{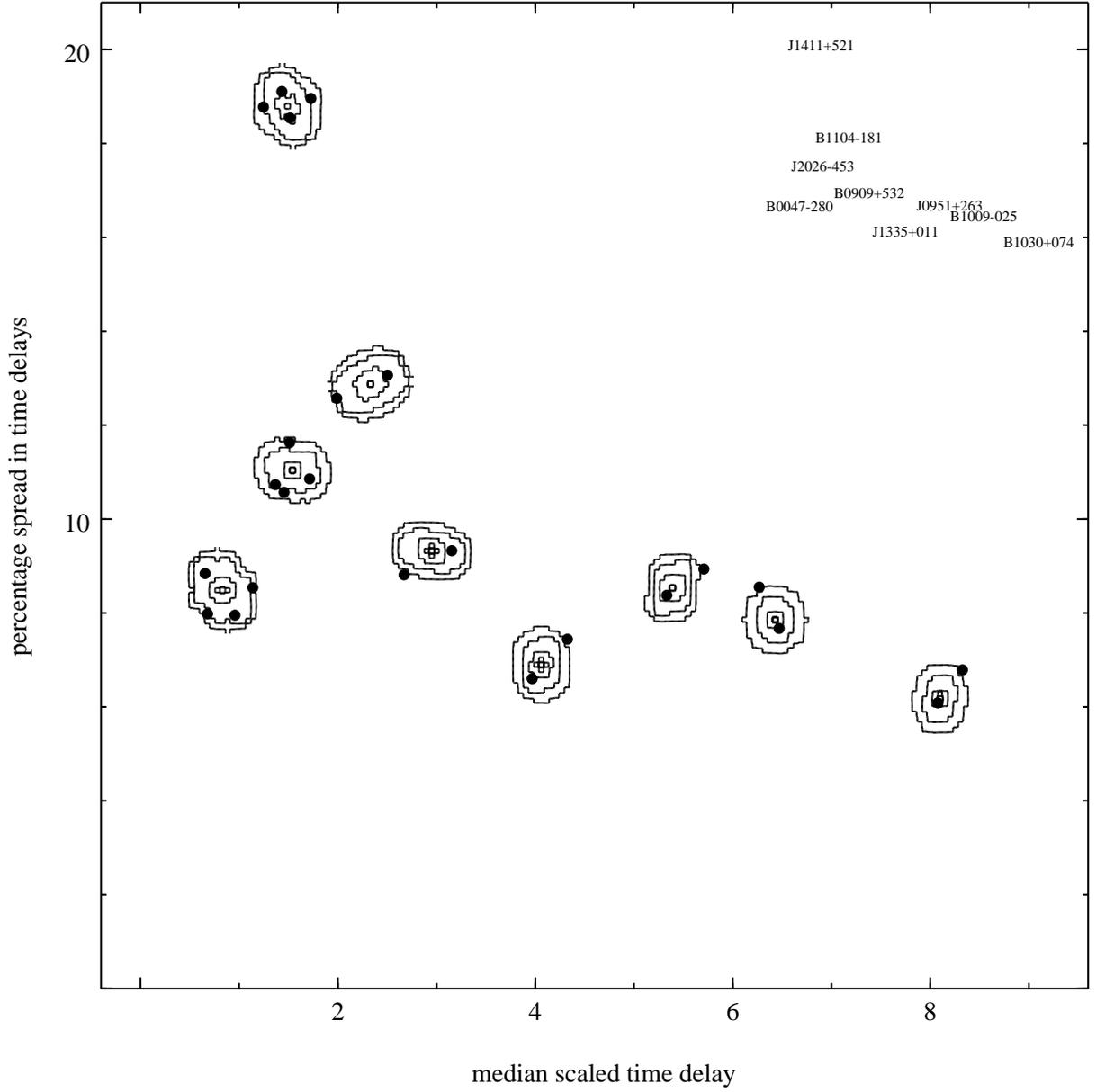}
\caption{Continuation of Figure~\ref{map1}. All models have inversion
symmetry.}
\label{map2}
\end{figure}

\begin{figure}
\epsscale{1}
\plotone{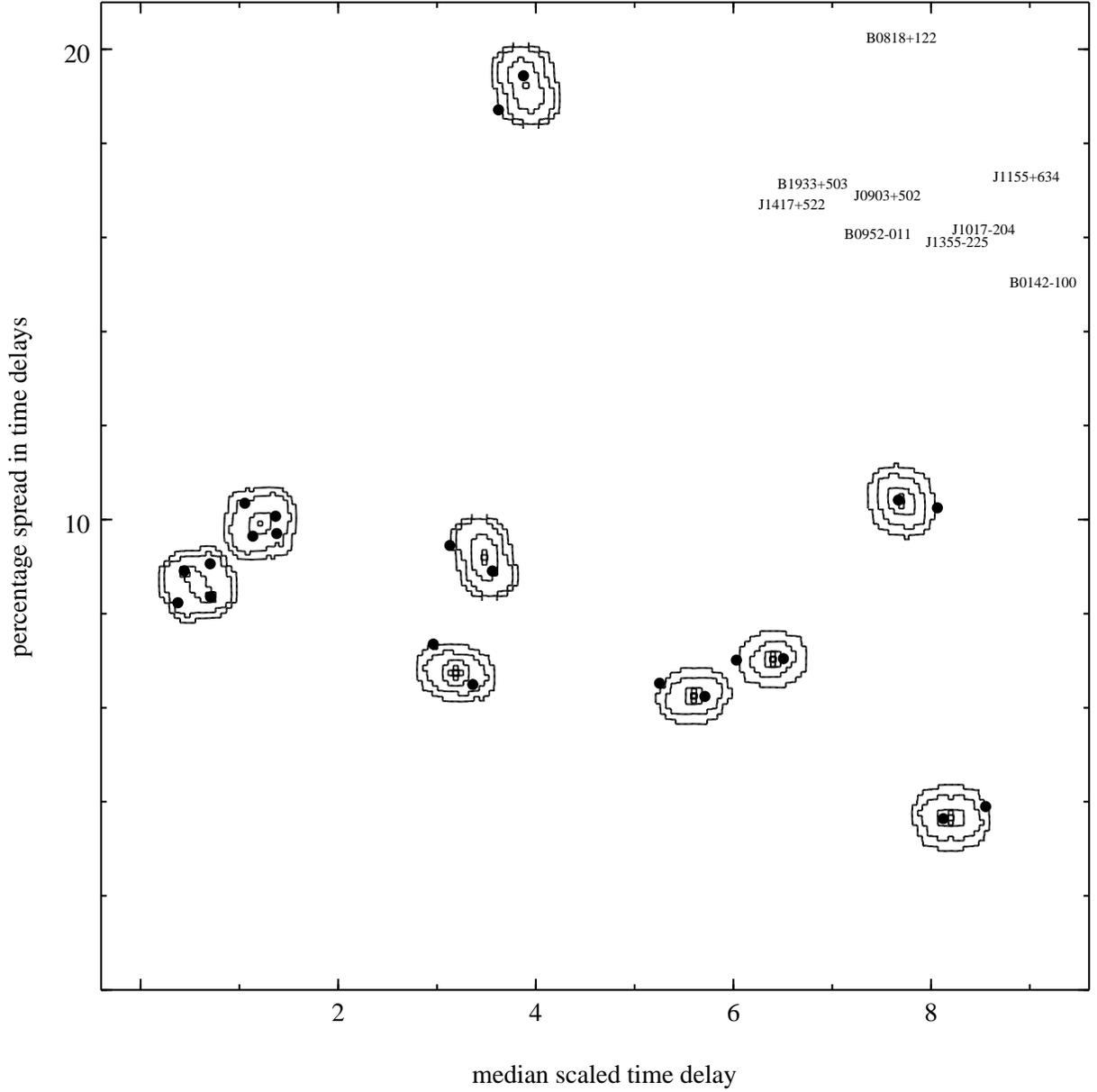}
\caption{Continuation of Figures~\ref{map1} and \ref{map2}. In the
case of the ten-image system B1933+503, we used all images for
modeling, but considered $\varphi$ for the core quad, as
indicated. All models have inversion symmetry.}
\label{map3}
\end{figure}

\begin{figure}
\epsscale{1}
\plotone{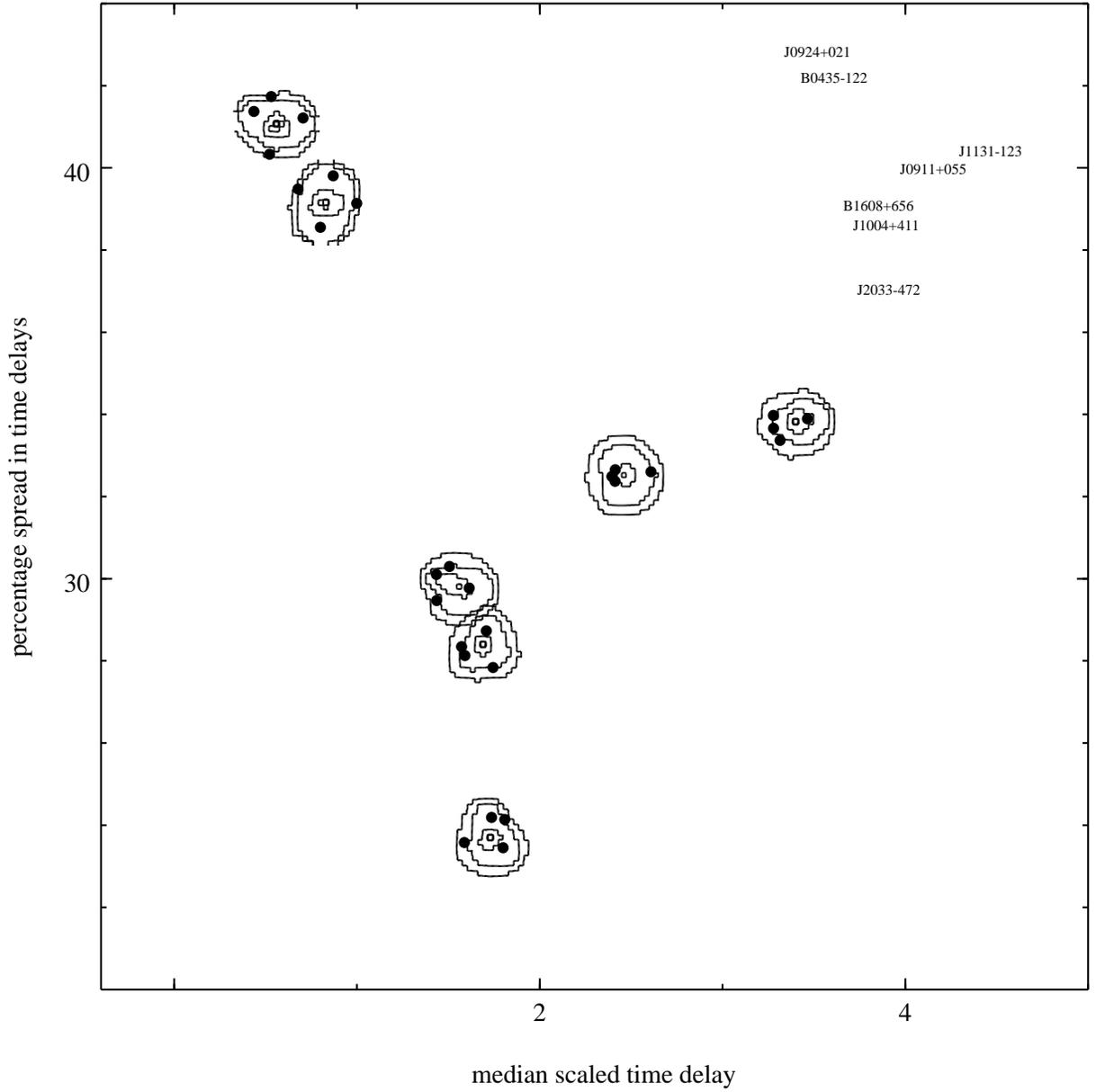}
\caption{Similar to Figs.~\ref{map1}--\ref{map3}, but using a
different scale.  All the lenses are asymmetric with very large
$\Delta\varphi/\varphi$.}
\label{map4}
\end{figure}

\begin{figure}
\epsscale{.25}
\plotone{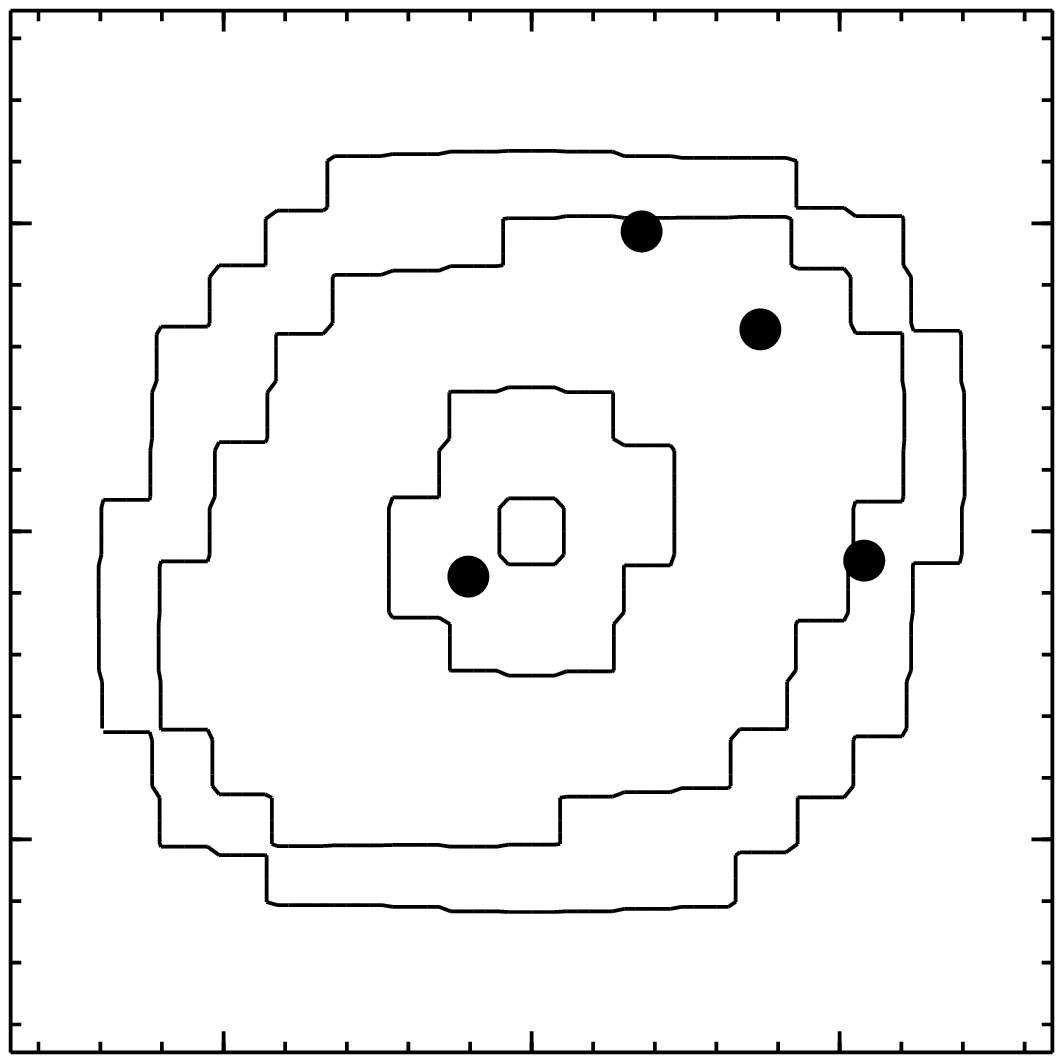} \goodbreak
\plotone{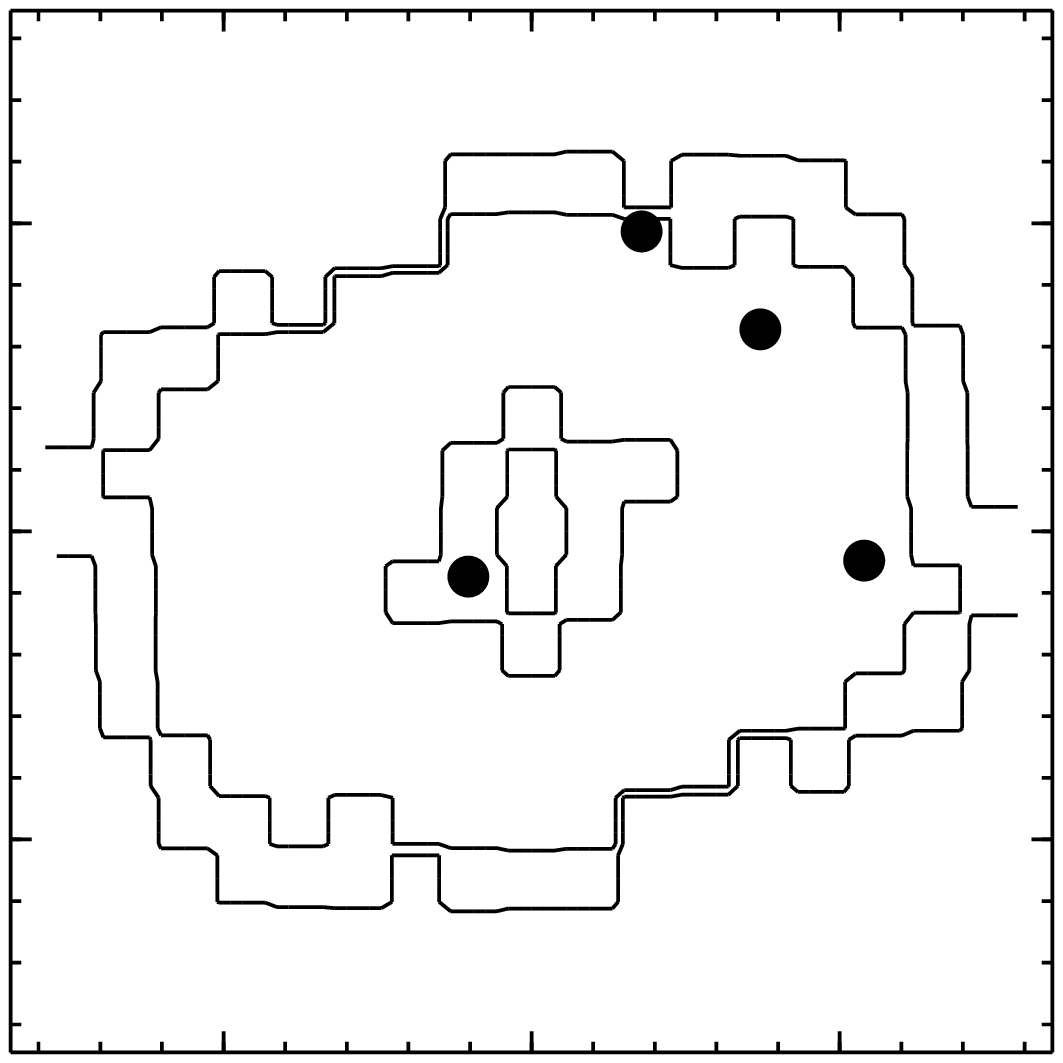} \goodbreak
\plotone{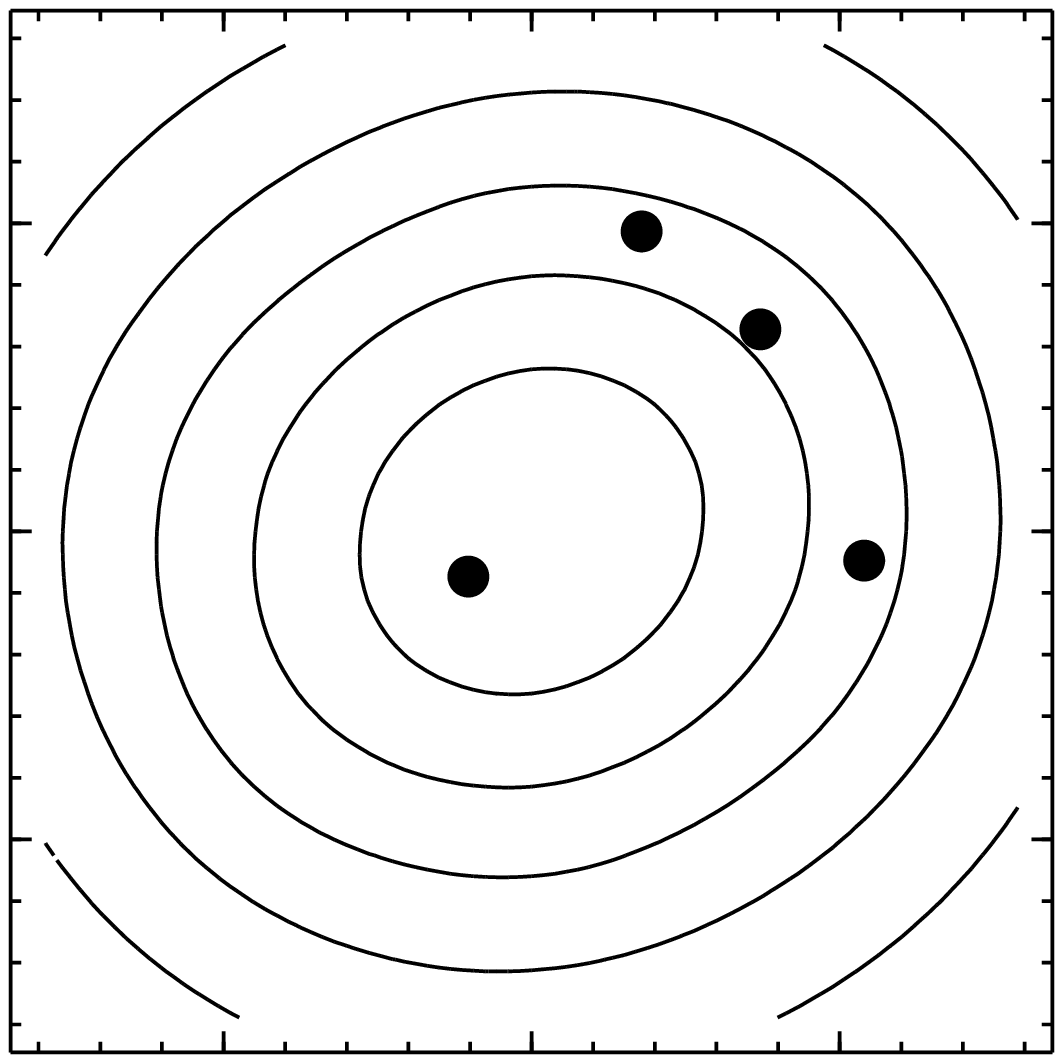} \goodbreak
\plotone{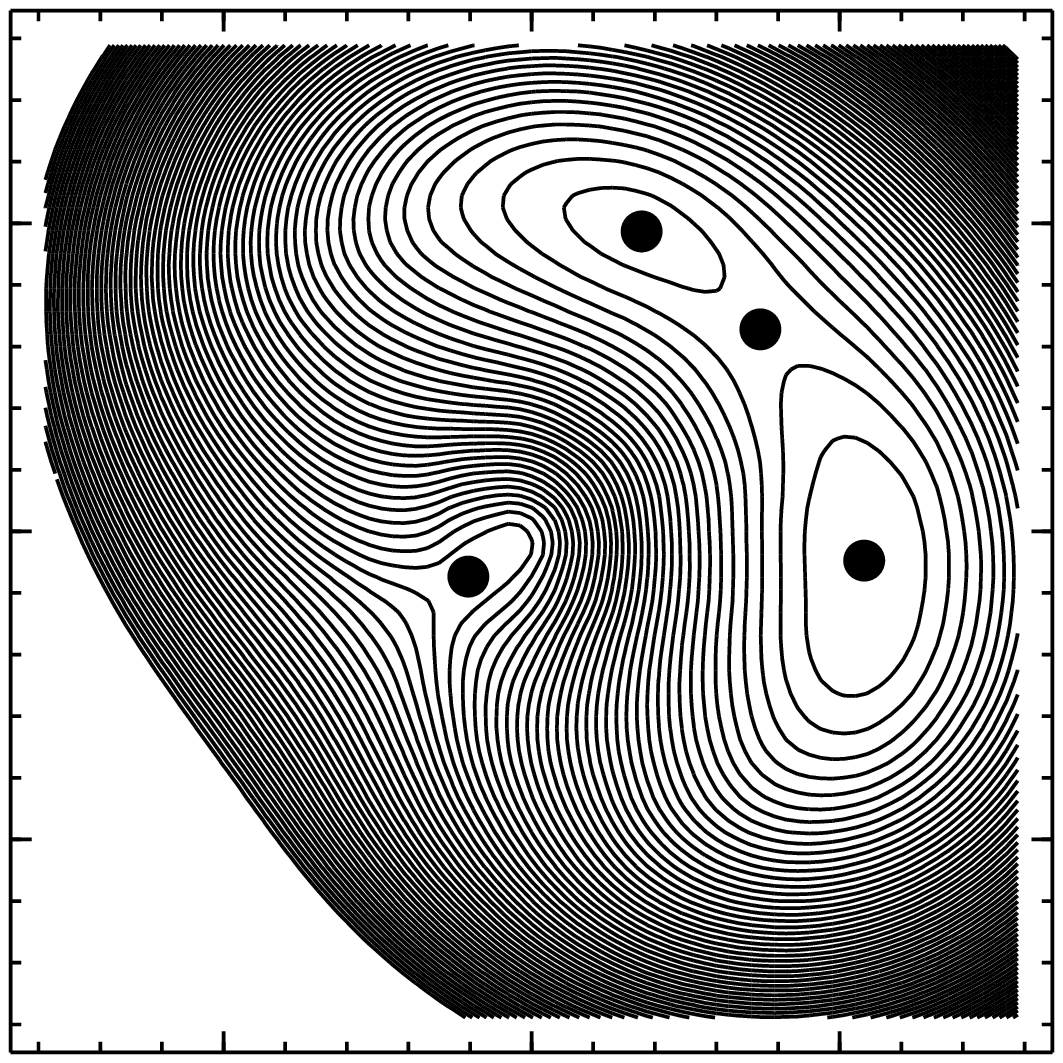}
\caption{Models for B1422+231.  Top panel:~ensemble-average mass map,
a miniature of which appears in Fig.~\ref{map1}. Second panel:~mass
map of a randomly chosen sample model from the ensemble; note the
larger pixel-to-pixel variation.  Third panel:~lens potential for the
galaxy in the sample model (external shear potential omitted);
we see here how solving for the potential automatically smoothes
out small-scale fluctuations in the mass. Bottom
panel:~arrival-time surface for the sample model; note that no
spurious extra images are present.}
\label{isol1}
\end{figure}

\begin{figure}
\epsscale{.25}
\plotone{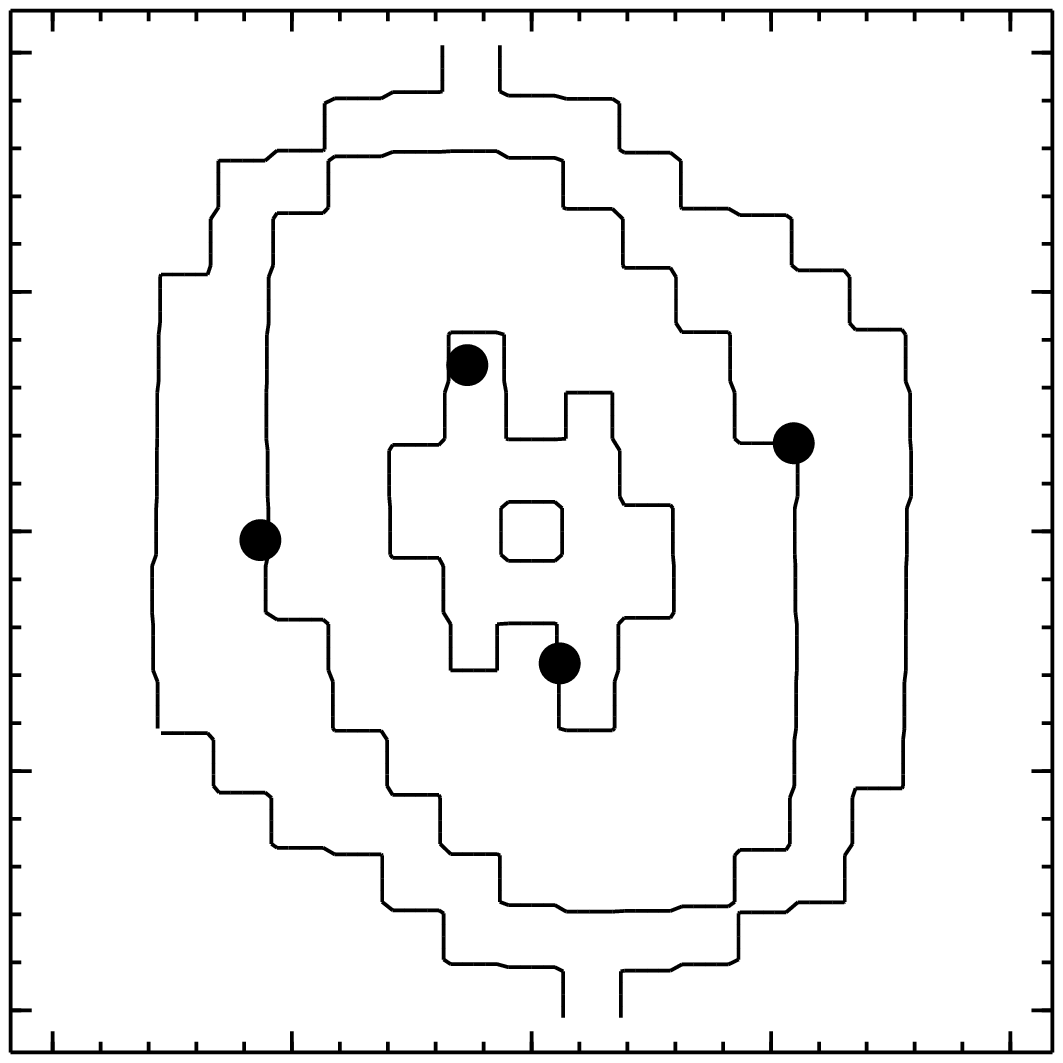} \goodbreak
\plotone{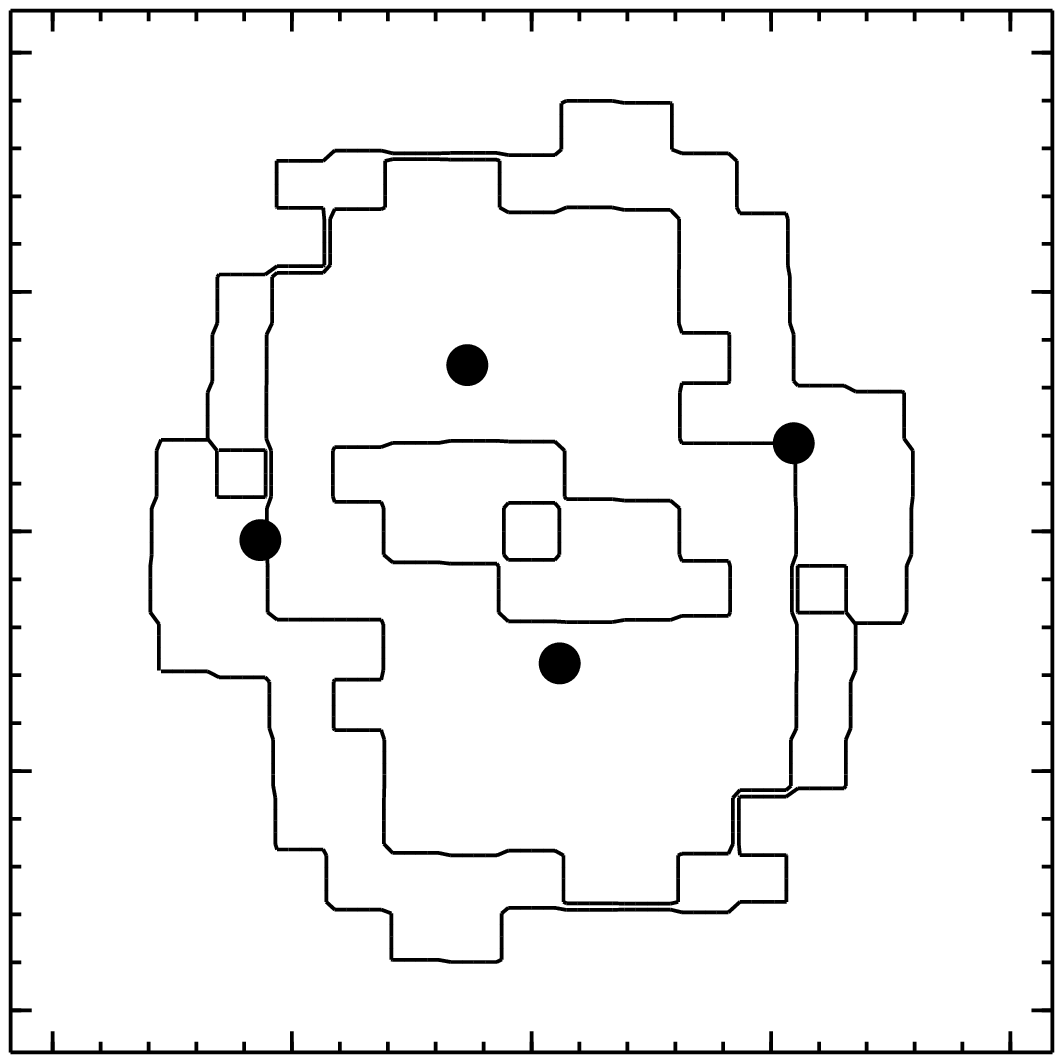} \goodbreak
\plotone{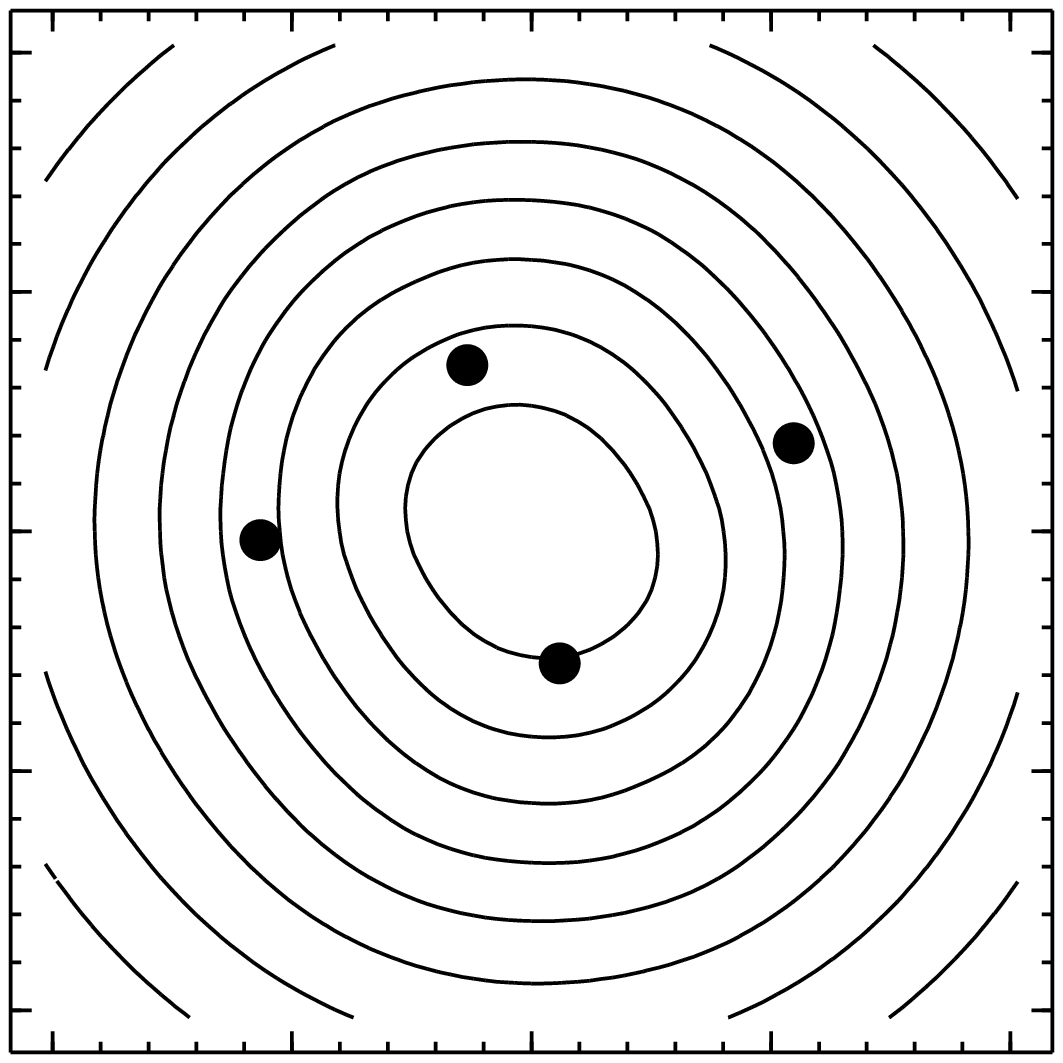} \goodbreak
\plotone{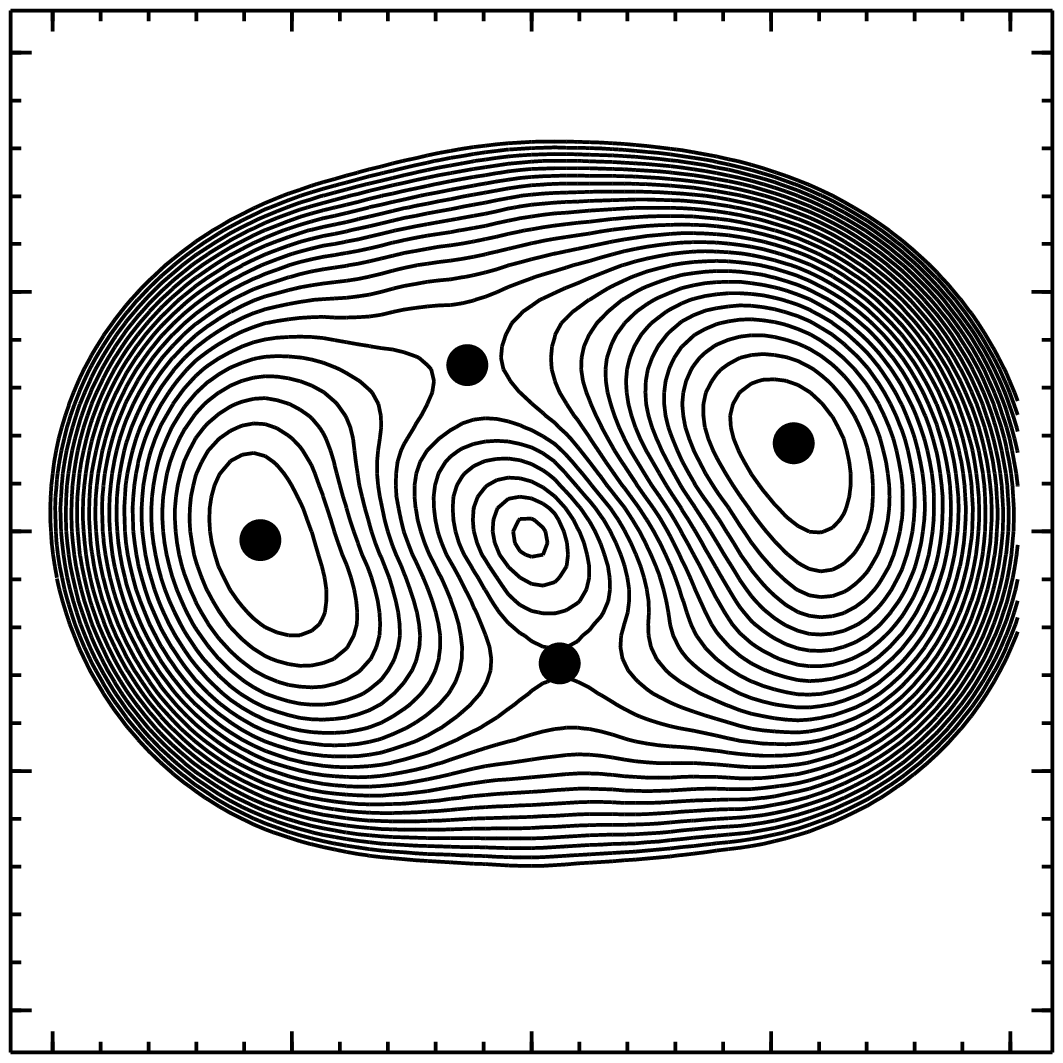}
\caption{Like Fig.~\ref{isol1} but for J1411+521. The top panel
appears in miniature in Fig.~\ref{map2}. The sample model (middle panel)
has a bar, while the ensemble as a whole does not.}
\label{isol2}
\end{figure}

\end{document}